\documentclass[lettersize,journal]{IEEEtran}
\usepackage{cite}

\usepackage{amsmath,amssymb,amsfonts}
\usepackage{algorithmic}
\usepackage{graphicx}
\usepackage{textcomp}
\usepackage{xcolor}
\usepackage{tikz}
\usepackage{pgf}
\usepackage{url}
\usepackage{multirow}
\usetikzlibrary{arrows,positioning,matrix,scopes,fit,automata}

\usepackage{listings}

\definecolor{codegreen}{rgb}{0,0.6,0}
\definecolor{codegray}{rgb}{0.5,0.5,0.5}
\definecolor{codepurple}{rgb}{0.58,0,0.82}
\definecolor{backcolour}{rgb}{0.95,0.95,0.92}

\lstdefinestyle{mystyle}{
    commentstyle=\color{codegreen},
    keywordstyle=\color{magenta},
    basicstyle=\ttfamily\footnotesize,
    breakatwhitespace=false,         
    breaklines=true,                 
    keepspaces=true,                 
    showspaces=false,                
    showstringspaces=false,
    showtabs=false,                  
    tabsize=2
}

\newcommand{\squishlist}{
 \begin{list}{$\bullet$}
   { \setlength{\itemsep}{0pt}
     \setlength{\parsep}{0pt}
     \setlength{\topsep}{0pt}
     \setlength{\partopsep}{0pt}
     \setlength{\leftmargin}{2.5em}
     \setlength{\labelwidth}{1.5em}
     \setlength{\labelsep}{0.5em} } } 

\newcommand{\squishend}{
  \end{list}  }
  
\def\BibTeX{{\rm B\kern-.05em{\sc i\kern-.025em b}\kern-.08em
    T\kern-.1667em\lower.7ex\hbox{E}\kern-.125emX}}
\begin{document}

\title{\textsc{ComCat}: Leveraging Human Judgment to Improve Automatic Documentation and Summarization}

\author{
    Skyler Grandel, Scott Thomas Andersen, Yu Huang, and Kevin Leach

    \thanks{Skyler Grandel, Yu Huang, and Kevin Leach are with the Department of Computer Science, Vanderbilt University, Nashville, Tennessee, USA. Email: \{skyler.h.grandel, yu.huang, kevin.leach\}@vanderbilt.edu.}
    \thanks{Scott Thomas Andersen is with the Department of Computer Science, Universidad Nacional Aut\`onoma de M\`exico, Mexico City, Mexico. Email: stasen@comunidad.unam.mx}

}

\maketitle

\begin{abstract}

Software maintenance constitutes a substantial portion of the total lifetime costs of software, with a significant portion attributed to code comprehension. 
Software comprehension is eased by documentation such as comments that summarize and explain code. 
We present \textsc{ComCat}, an approach to automate comment generation by augmenting Large Language Models (LLMs) with expertise-guided context to target the annotation of source code with comments that improve comprehension.
Our approach enables the selection of the most relevant and informative comments for a given snippet or file containing source code. 
We develop the \textsc{ComCat} pipeline to comment C/C++ files by (1) automatically identifying suitable locations in which to place comments, (2) predicting the most helpful type of comment for each location, and (3) generating a comment based on the selected location and comment type.  
In a human subject evaluation, we demonstrate that \textsc{ComCat}-generated comments significantly improve developer code comprehension across three indicative software engineering tasks by up to 12\% for 87\% of participants.
In addition, we demonstrate that \textsc{ComCat}-generated comments are at least as accurate and readable as human-generated comments and are preferred over standard ChatGPT-generated comments for up to 92\% of snippets of code.
Furthermore, we develop and release a dataset containing source code snippets, human-written comments, and human-annotated comment categories.
\textsc{ComCat} leverages LLMs to offer a significant improvement in code comprehension across a variety of human software engineering tasks.

\end{abstract}

\begin{figure*}
    \centering
        \includegraphics[width=\linewidth]{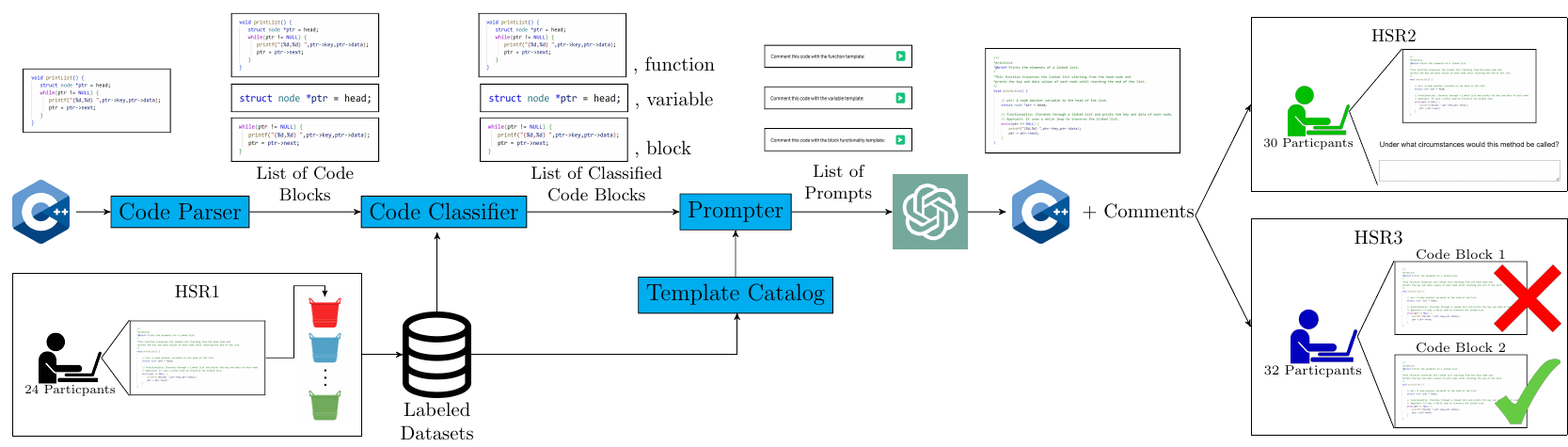}
    \vspace{-4ex}
    \caption{\textsc{ComCat} pipeline and study procedure. We use three instances of human subjects research to inform \textsc{ComCat}'s design (1) and evaluate developer performance (2) and preference (3) with our tool. \textsc{ComCat} takes C/C++ code as input, using a Code Parser to identify code Snippets to be commented. These Snippets are classified, and the class of each Snippet is used in combination with our Template Catalog to create a prompt for each Snippet. These prompt ChatGPT, which outputs the commented code. This pipeline is informed by developer expertise, but it is fully automated and requires no human intervention.}
    \vspace{-1ex}
    \label{fig:studies}
    \vspace{-2ex}
\end{figure*}

\section{Introduction}
\label{sec:intro}
Software maintenance accounts for a significant portion of the total lifetime costs of software, with estimates ranging from 66\% to 90\%~\cite{erlikh_2000}. 
Within the realm of maintenance, approximately half of the costs are attributed to code comprehension~\cite{foster_1993, nguyen_2010, yip_1994}. 
Furthermore, a substantial portion of the time spent on comprehension is dedicated to reading plain code, comprising over 40\% of the overall comprehension time~\cite{latoza_2006}. 
Given these statistics, it is evident that software readability and comprehension play pivotal roles in the cost-effectiveness and efficiency of software development and maintenance processes. 
To address this critical aspect of the software life cycle, we present a method to improve automated software documentation by augmenting Large Language Models (LLMs) with an awareness of (1) categories and (2) appropriate locations of source code comments that are identified as helpful according to human software developers. 

The importance of comments in code comprehension has been a topic of debate among
researchers~\cite{aghajani2020software, borstler_2016}. While B\"{o}rstler and Paech argue that comments primarily enhance perceived readability rather than objectively improve comprehension~\cite{borstler_2016}, other studies indicate that commented code is easier to understand~\cite{stapleton2020human,tenny_1988, woodfield_1981} and comments are particularly useful for debugging, comprehending, and refactoring~\cite{aghajani2020software, stapleton2020human}. 
Moreover, inadequate comments can substantially impede software maintainability~\cite{lientz_1983}. 
These contrasting viewpoints imply that, while certain comments may not contribute to improved comprehension, the appropriate placement of complete, concise, and accurate comments can greatly enhance the efficiency of code reading. Consequently, the identification, placement, and generation of comments that can improve comprehension emerges as a crucial problem to address.

Automated code documentation generation, or code summarization, has seen several advancements in recent years with the goal of supplanting or supplementing manual documentation~\cite{abid2015using,ahmad2021unified,allamanis2016convolutional,barone2017parallel,devlin2018bert,eddy2013evaluating,feng2020codebert,gao2023code,haiduc2010supporting,haiduc2010use,hu2018deep,hu2020deep,iyer2016summarizing,khan2022automatic,liu2020retrieval,mcburney2014automatic,moreno2013automatic,parvez2021retrieval,phan2021cotext,rastkar2011generating,rodeghero2014improving,sridhara2010towards,wan2018improving,wang2021codet5,wong2013autocomment,wong2015clocom}. 
Researchers have improved performance in this task using context-aware techniques, and text classification in particular has been shown to improve performance for multi-intent comment generation~\cite{chen2021my, geng2024large}.
While previous research focuses on generating comments for function-level summaries alone, our work recognizes the critical need for inline comments within functions in a variety of contexts.
Such comments play a pivotal role in clarifying complex code logic on a granular level, 
which has been largely overlooked as a means for enhancing comprehensibility.
In this paper, we address two key challenges:
(1) identifying the types and locations of comments that are most useful for developers; and 
(2) demonstrating improved comprehension by human developers without relying on 
traditional evaluation metrics like BLEU, which do not correlate well with developer performance~\cite{stapleton2020human} and that may not be appropriate for recent developments in LLMs~\cite{song_2019,wang2023chatgpt}.
By leveraging human studies to guide context-aware comment generation, we enhance LLMs to meet these challenges for more granular and comprehensible documentation of source code.

In this paper, we present \textsc{ComCat}, an approach to automatically generate source code comments for several levels of abstraction, illustrated in Figure~\ref{fig:studies}.
Given an input source code file,
(1) the Code Parser splits the file into Snippets that capture commonly-used structures, like loops and variable declarations,
(2) the Code Classifier predicts the most helpful type of comment to place within each Snippet, and
(3) the Prompter uses an LLM to generate a comment for each Snippet and type of comment required.
\textsc{ComCat} outputs a well-documented source code file with each Snippet and corresponding comment.
We conduct three Human Subject Research (HSR) studies in this paper\footnote{All HSR in this paper was exempted by our institution's IRB.}. 
First, we surveyed 24 developers (HSR1 in Figure~\ref{fig:studies}) to determine which types of comments are most helpful for comprehension, from which we trained the Code Classifier and built a Template Catalog for prompting LLMs for comments.
Next, we recruited 30 additional developers (HSR2 in Figure~\ref{fig:studies}) to evaluate how much \textsc{ComCat} improved developer comprehension in three different software engineering tasks.
Finally, to determine the effectiveness of our pipeline, we recruited 32 different developers (HSR3 in Figure~\ref{fig:studies}) to determine whether developers preferred \textsc{ComCat} over human-written and ChatGPT-generated comments.
We show that our approach significantly improves developer comprehension and that developers significantly prefer \textsc{ComCat} over other approaches to generating source code comments and summaries.
Finally, we release our dataset of source code, snippets, human-written comments, and anonymized human subject data for future research. 

\noindent The main contributions\footnote{Our replication package can be found at: \\\url{https://osf.io/tf2eu/?view_only=4dcf8efa50a346dc96d28b139b0d3b90}.} of this paper are:
\squishlist
\item A set of developer-informed comment categories for comment analysis with a corresponding classifier and dataset of comments and their associated C/C++ code annotated via a HSR study with 24 participants.
\item A catalog of prompts for LLMs to produce comments of each relevant comment class.
\item A tool for dynamically selecting and automatically commenting code Snippets in a C/C++ file using expertise-guided classification and prompting with an LLM. 
\item A second human study of 30 participants demonstrating that our context-generation improves developer performance on code comprehension tasks by an average of 12\% for 87\% of developers.
\item A third HSR study of 32 participants demonstrating that our context-generation leads to documentation that is at least as accurate and readable as what was generated by humans in 80 to 100\% of cases and is preferred to standard ChatGPT in 60 to 92\% of cases.
\squishend

\section{Background and Related Work}\label{sec:background}

Source code documentation is an active area of research that we build upon by improving analysis and generation of software comments. In this section, we consider two lines of related work: (1) analysis and categorization of software documentation, and (2) techniques for automatically documenting source code. We acknowledge the contributions of these works and leverage their insights to better generate comments.

\subsection{Code Documentation Quality Analysis and Categorization}

Other researchers have conducted a number of studies on comment quality evaluation.
The most similar to our work is a study by Steidl et al.~\cite{steidl_2013}, which produces a quality assessment model and a set of systematic metrics for evaluation.
They further classify comments, largely by the code they annotate, into seven categories: copyright comments, header comments, member comments, inline comments, section comments, code comments, and task comments.
This work led to a machine learning model that automatically classifies comments into these categories.
A similar work by Chen et al.~\cite{chen2021my} uses manual categorization by purpose into six categories to improve code summarization in a similar way to our work, though they only consider comments that describe entire methods.
In this paper, we re-examine automatic comment classification as a tool for improving comment generation and assessing comment quality using a more detailed categorization schema that better distinguishes inline comment types.

Khamis et al.~\cite{khamis_2010} present JavadocMiner, a tool for automatically calculating the language quality and cohesion with code of Javadoc comments. It additionally analyzes an entire project to calculate how well commented it is. This study produces a number of metrics and heuristics to determine how well a function comment documents a method and how well a file or project is documented overall. To the best of our knowledge, we are unaware of any previous attempted use of HSR for comment classification or any attempts at experimentally evaluating the quality of comments. We address this gap by investigating the quality of comments generated by LLMs and how it can be improved by Transformer-based classification.

\subsection{Automated Code Documentation}

Automatic documentation typically involves three approaches: 
(1) template-based, (2) information retrieval, and (3) learning-based.
We discuss each below.

\noindent\textbf{Template-based} tools use a predefined layout with rules to produce documentation. Some have used natural language templates to capture details of functions~\cite{sridhara2010towards} while others have used heuristics~\cite{moreno2013automatic, rastkar2011generating}, contextually informed templates~\cite{mcburney2014automatic}, and stereotyping with a code analysis framework~\cite{abid2015using}. Buse and Weimer further used templates to document code changes using symbolic exectution~\cite{buse2010automatically} and exceptions~\cite{buse2008automatic}.

\noindent\textbf{Information Retrieval} tools seek out existing documentation to copy into new code. Haiduc et al. used latent semantic indexing and vector space modeling to generate comments for functions and classes~\cite{haiduc2010supporting, haiduc2010use}. Eddy et al. contributed a hierarchical model~\cite{eddy2013evaluating} while Rodeghero et al. improved upon it by using eye-tracking data to inform statement and keyword selection~\cite{rodeghero2014improving}. Code clone detection has also been leveraged to mine StackOverflow for comments by  Wong et al.~\cite{wong2015clocom, wong2013autocomment}.

\noindent\textbf{Learning-based} tools typically use deep learning techniques to learn latent features from source code by training on existing code/summary pairs. CODE-NN, an LSTM-based neural network by Iyer et al., was trained on StackOverflow code to summarize C\# and SQL code~\cite{iyer2016summarizing}. Convolutional Neural Networks~\cite{allamanis2016convolutional} and Neural Machine Translation~\cite{barone2017parallel} have also been used for code summarization. Hybrid approaches combining code with ASTs~\cite{wan2018improving} or lexical and syntactical information~\cite{hu2018deep, hu2020deep} have also been popular. 

Other works have used Transformer-based methods.
CodeBERT~\cite{feng2020codebert} and CodeT5~\cite{wang2021codet5} are pre-trained transformer models based on BERT~\cite{devlin2018bert} and  T5~\cite{raffel2020exploring} respectively.
GPT-2, CodeGPT-2, and CodeGPT-adapted~\cite{radford2019language, ahmad2021unified} have all also been evaluated on this task using the decoder-only GPT-2 base model.
The Code Structure Guided Transformer ~\cite{gao2023code} incorporates code structural properties into the transformer to improve summarization performance.
CoTexT~\cite{phan2021cotext} learns the representative context between natural and programming languages and performs several downstream tasks, including code summarization.
PLBART~\cite{ahmad2021unified} is a sequence-to-sequence model pre-trained on a large set of Java and Python functions and their textual descriptions.
Liu et al.~\cite{liu2020retrieval} uses a retrieval-augmented GNN to generate comments for C/C++ source code. 
Hybrid approaches have also been used, such as incorporating relevant codes/summaries retrieved from a database~\cite{parvez2021retrieval}.
Others have used LLMs like Codex, a GPT-3 based tool evaluated on summarization by Kahn and Uddin~\cite{khan2022automatic}.
Similar to our work is a study by Geng et al.~\cite{geng2024large}, who combine Codex with a taxonomy of comments from Chen et al.~\cite{chen2021my} to produce multiple summaries from different perspectives. We build on this by extending the taxonomy to enable inline comment generation, introducing dynamic code Snippet selection, and evaluating human perceptions of our resulting comments.
CodeBERT, CodeT5, GPT-2, CodeGPT-2, CodeGPT-adapted, CoTexT, and Codex were all trained and/or evaluated using CodeSearchNet~\cite{husain2019codesearchnet}, a dataset of functions and their associated documentation in six languages.

Our approach dynamically selects code Snippets and corresponding comment types to generate the most helpful comment as informed by human developers. 
Our approach can apply both to inline Snippets of code as well as entire functions~\cite{khamis_2010, sridhara2010towards}.
To the best of our knowledge, our work is the first to generate both types of comments enabled by comment type classification.

\section{Comment Classification Approach}
\label{classification}
A critical foundation of our approach is to identify the best type of comment to place in source code as well as where to place each comment. 
In this section, we present our approach for classifying comments, which we use in our overall pipeline to provide targeted, comprehensible comments in source code. 
This work enables developers to thoroughly and accurately document code with software comments automatically, saving substantial monetary and temporal costs associated with software development and maintenance~\cite{erlikh_2000, foster_1993, latoza_2006, nguyen_2010, yip_1994}. 
With this goal in mind, the following subsections detail our HSR-based classification and comment templates. 
These components form the basis for our expert-guided context generation to be used in our \textsc{ComCat} pipeline for fine-grained software documentation. 

The code and comments in this study (and in all HSR studies in this paper) were scraped from C/C++ GitHub projects originating from the DARPA MUSE project~\cite{ellingwood2019draper}.  
This dataset was formatted with Clang to ensure a consistent style~\cite{clang}.

\subsection{Classification Schema}
\label{categories}
To establish appropriate comment types for this work, we undertake a systematic process of comment categorization based on the information comments provide about their associated code. 
We build on previous works by identifying distinct comment types that capture the semantics and purpose of each comment, with the end goal of providing more descriptive inline comment types than previous works to encapsulate the varied information and structures of such comments~\cite{chen2021my, steidl_2013}. 
To ensure the accuracy and generalizability of this process, two researchers independently categorize comments, starting with zero categories and creating a new category when no existing category applies to the current comment. Following accepted qualitative methods~\cite{hennink2017code, miner2006saturation}, the categorization process continues until saturation\footnote{In this study, \emph{saturation} refers to the point where no new categories have been added to the codebook after at least 100 consecutive comments have been categorized without creating a new type.} is reached, ensuring comprehensive coverage of comment types.

Following this process, we take the union of the individual categorizations of the two researchers, and equivalent categories are consolidated to form a comprehensive schema through discussions and iterative refinement, resulting in a schema of 12 comment types. 
To further validate this schema, we conducted HSR1 (described in Section~\ref{hsr1}). 
HSR1 revealed four of the initial 12 types are perceived to be useful for improving developer understanding of source code, finalizing our schema with the following classes:

\squishlist
\item \textbf{Function} comments describe an entire function. They tend to note parameters and return values.
\item \textbf{Variable} comments describe a variable, constant, or literal. They often note what a variable represents.
\item \textbf{Snippet Functionality} comments are inline comments that summarize or describe the functionality of code.
\item \textbf{Branch} comments describe possible branches of execution. This also includes preconditions for branches.
%\item \textbf{Reasoning} comments describe the reasoning behind implementation decisions, but not functionality.
%\item \textbf{Quirk} comments contain a random quirk of the code, author jokes, or some other unimportant information.
%\item \textbf{Use Guideline} comments guide readers on using or accessing functions, containers, or variables, or they detail compilation or execution instructions.
%\item \textbf{Source} comments describe the source of the code. These might note that the code was copied from some link.
%\item \textbf{Copyright} comments contain copyright, licensing, and author information.
%\item \textbf{Section} comments provide a section label for multiple functions, test cases, or global or class variables. 
%\item \textbf{Code} comments are commented out code.
%\item \textbf{Task} comments note future work, e.g. a TODO/FIXME. 
\squishend

\subsection{HSR1: A Human Study of Comment Classification}
\label{hsr1}
To categorize our dataset of comments, we conduct HSR1, a survey of 24 student and professional developers.
We recruit participants from both student and professional populations of developers with at least 2 years of C/C++ experience from various universities and companies through email solicitations and online crowdsourcing\footnote{We specifically gather 10 participants in HSR1 from Mechanical Turk and 15 participants in HSR2 from Prolific for crowdsourcing. All other participants were student volunteers from various universities.}. 
We employ a C/C++ prescreening test to ensure adequate programming aptitude.
This prescreening consists of questions found by Sillito et al. to be relevant to developer comprehension, similar to the questions employed by our study in HSR2~\cite{sillito_2006}.
The survey is administered online via a URL given to participants. 
Participants are provided with a training video giving standard and edge-case examples of comments and their associated categories, along with an explanation of each categorization. 

In HSR1, we ask participants to classify comments drawn randomly from our dataset according to the categories from our schema.
Participants also have an option to note that a given comment does not fit any category.
We further ask participants to indicate whether the comment aids in their understanding of the code. 
Participants are given the entirety of a file for context and indicate a ranked choice of up to three categories for a given comment to resolve disagreements and to allow for comments that exhibit qualities of multiple types. 
We also ensure that at least two participants classified every example in the dataset.
Although participants could report new comment types if they chose, we found that such cases consisted of existing types. 

We further calculate the inter-rater reliability of the results using Krippendorf's Alpha ($\alpha$)~\cite{gwet2011krippendorff}. 
This statistic is a generalization of Fleiss' Kappa to inconsistent numbers of respondents~\cite{fleiss1971measuring}.
This results in an $\alpha$ of 0.56 across 24 participants and 12 nominal categories. 
This can be interpreted as a ``moderate agreement'' for a task where chance agreement is unlikely due to the large number of categories. 
While this indicates reliable data, 
examples can exhibit characteristics of multiple classes, resulting in occasional disagreement. 
We resolve such disagreements by selecting the ranked choice majority category to be the accepted class and resolving any remaining disagreements through discussion among two authors.
HSR1 resulted in a dataset of 952 examples of comments and associated human-annotated comment categories. 
These comment categories were ultimately used for training a comment classifier used in subsequent parts of our pipeline.

Having established a schema of comment categories, we next define a set of comment templates. 
Recall that our eventual goal is to categorize the type and location in which to place comments in source code.
Each comment type is associated with a template prompt used to prompt an LLM to generate a comment of that type in a specific context. 
We use developer expertise to guide the comment type that we use to describe a given code Snippet. 
However, we found that not every type of comment is useful for the specific purpose of comprehending code.
Therefore, we constrain our comment types to include only those that developers in HSR1 found useful for comprehension in a majority of cases. 
Thus, we create templates based on the structure of comment type that was considered useful for comprehension in HSR1, and discuss each template below.

\begin{table*}[t]
\caption{Our Template Catalog. This catalog associates our comment types with a template for producing a comment of that type. Text in brackets should be replaced with information specific to the code being commented.\vspace{-9pt}}
\label{table:catalog}
\begin{tabular}{ |p{3cm}|p{14cm}| }
 \hline
 \multicolumn{2}{|c|}{Template Catalog} \\
 \hline
 \multirow{12}{*}{\textbf{Function}} & 
 \vspace{-6pt}
 \begin{lstlisting}[language=C]
/**
 * [Function name]
 * @brief [Brief description of the function]
 *
 * [Provide a detailed explanation of the function's purpose and functionality]
 *
 * @param [param1] [Description of the first parameter]
 * @param [param2] [Description of the second parameter]
 * @return [Description of the return value]
 *
 * @exception [ExceptionType] [Description of the exception and when it can occur]
 */
 \end{lstlisting}
 \vspace{-18pt}
 \\
 \hline
 \multirow{2}{*}{\textbf{Variable}} & 
 \vspace{-6pt}
 \begin{lstlisting}[language=C]
// [Variable/Constant/Literal Name]: [Description and purpose].
 \end{lstlisting}
 \vspace{-18pt}
 \\
 \hline
 \multirow{3}{*}{\textbf{Snippet Functionality}} & 
 \vspace{-6pt}
 \begin{lstlisting}[language=C]
// Function: [Functionality and purpose].
// Approach: [Approach or strategy].
 \end{lstlisting}
 \vspace{-18pt}
 \\
 \hline
 \multirow{3}{*}{\textbf{Branch}} & 
 \vspace{-7pt}
 \begin{lstlisting}[language=C]
// Cond: [Conditions]
// Function: [Functionality and Purpose]
 \end{lstlisting}
 \vspace{-18pt}
 \\
 \hline
\end{tabular}
\vspace{-4ex}
\end{table*}

\subsection{Template-Based Prompting}
\label{prompting}

Recall that our goal is to automatically document source code based on the best predicted comment type and location.  Having already established a developer-validated schema of useful comment types, 
we next design prompt templates for LLMs, which we treat as black box state-of-the-art comment generators~\cite{khan2022automatic}. 
By augmenting the prompt with a predicted location and helpful comment type, we can target the specific type of comments for the LLM to generate that are more likely to be helpful for code comprehension according to HSR1.  

In Section~\ref{hsr1}, we developed a scheme of 12 comment types, and asked participants to annotate which comment types they believed were most helpful for comprehension. 
Thus, for building our prompt templates, we consider only the comment categories that were judged to be useful by a majority of participants in HSR1. 
We further acknowledge that practitioners prefer documentation that explains what the associated code is doing as well as why~\cite{hu2022practitioners}. 
Therefore, we consolidate ``Reasoning'' into our 4 templates by prompting for the code's purpose so that all of our documentation explains both the ``what'' and the ``why'' for some given code. 
We further note that some developers consider ``Use Guideline'' comments to be useful~\cite{geng2024large}, so we include use-guideline elements in our ``Function'' template.

To ensure that our LLM (i.e., ChatGPT-3.5) produces the comments in the format we desire, we use the Template prompt pattern~\cite{white2023prompt}. 
To develop our templates, we use common structures of each comment type in our dataset and the Flipped Interaction pattern~\cite{white2023prompt} in combination with testing ChatGPT for its adherence to the prompt, output length, and comment cohesion with the input code. 
These trials resulted in templates for function, variable, Snippet functionality, and branch comments, which can be found in Table~\ref{table:catalog}. 
All templates incorporate the reasoning type to explain implementation decisions for all input code by prompting for the code's purpose. 

The templates in Table~\ref{table:catalog} are included in a prompt that instructs ChatGPT to generate a comment for the given code Snippet using the template. We further use the OpenAI API system prompt to indicate ChatGPT's role as a comment generator, provide the entirety of the file for context, and limit the length of the output comments. 
We ultimately use these templates and their associated prompts to produce a targeted comment with content and structure that fits the semantics of a particular code Snippet.

\section{\textsc{ComCat} Pipeline Approach}
\label{sec:pipeline}
In this section, we present the \textsc{ComCat} pipeline architecture and detail its components. A visualization of this pipeline is given in Figure~\ref{fig:studies}. Taken as a whole, \textsc{ComCat} accepts a C/C++ file and outputs the same file with comments describing the functionality, purpose, and reasoning of the code embedded throughout.
\textsc{ComCat} fully automates the process of commenting C/C++ files to save costs associated with documenting code~\cite{erlikh_2000, foster_1993, latoza_2006, nguyen_2010, yip_1994} and conferring the benefits of well-documented code, like eased comprehension, debugging, refactoring, and reuse~\cite{aghajani2020software, stapleton2020human}. Quality is crucial to a comment's ability to confer these benefits and \textsc{ComCat}'s purpose is to ensure consistent, accurate, high-quality comments~\cite{lientz_1983, stapleton2020human, steidl_2013}. 
We accomplish this through careful preprocessing of input files to identify appropriate locations in which to generate developer-preferred comments, as well as the careful design of LLM prompts based on developer guidance.
While \textsc{ComCat} automatically comments files without the need for any human intervention, users may validate and edit the results to mitigate concerns about AI reliability.
In contrast to prior art, this approach strikes a balance between sufficient documentation at a granular level (as opposed to function-level summarization seen in prior work) and succinct documentation that is objectively helpful for developer comprehension (as opposed to AI approaches that are only internally validated with metrics like BLEU and ROUGE).

To achieve our goal, we 
(1) split an input source file into a number of code Snippets. Next,
we (2) determine an appropriate comment type to document each Snippet. Then, we
(3) select the most appropriate prompt template based on the selected comment type.
Next, 
we (4) fill the prompt template with contextual information from the code Snippet and type. 
Finally, we (5) feed each prompt to ChatGPT to build the final commented file. 
Therefore, this pipeline consists of five components: the Code Parser, Code Classifier, Template Catalog, Template-Based Prompter, and ChatGPT.
We discuss each below.

\subsection{Code Parser}
The Code Parser splits code into parts that ChatGPT will comment so that Snippets within the file can be documented. 
It takes C/C++ code as input and outputs a list of code Snippets to comment. 
Not every line from the source code will be included in the output, and code Snippets may overlap.
As shown in Figure~\ref{fig:studies}, 
code Snippets could be selected from various levels of abstraction, such as whole functions, single statements, or loop structures.

The Code Parser is implemented using the Clang C language family frontend for LLVM~\cite{clang}. 
This allows us to parse input code and identify constructs defined by several lines of code or several scopes within the code via an LLVM pass. 
This Code Parser selects every instance of the following C/C++ constructs defined by LLVM to be commented: functions/methods, conditionals, loops, try-catch-finally blocks, and declarations (e.g., of classes, structs, variables, etc.). 
We group associated Snippets together, so an if-else statement is considered a single Snippet. This functionality is provided entirely by LLVM. This component outputs a list of such Snippets to be passed on to the Code Classifier so that each code Snippet can be matched with a comment type (cf. Section~\ref{hsr1}) to describe it.

\subsection{Code Classifier}
\label{model}
The Code Classifier is a Transformer model that is trained to determine which of the four kinds of comment would be appropriate to describe the code Snippet. 
These four comment classes are (1) function comments, (2) variable comments, (3) Snippet functionality comments, and (4) branch comments (recall these were defined in our comment type schema in Table~\ref{table:catalog}). 
The Code Classifier takes a list of code Snippets from the Code Parser and outputs a list of classified Snippets. 
Each Snippet is passed to the prompter so that it can select the appropriate prompt to feed to ChatGPT to produce a comment for that Snippet. 

In this subsection, we describe the details of the architecture and training for the Code Classifier model. 
This is a multi-layer bidirectional Transformer model~\cite{vaswani2017attention} trained specifically for the task of classification with a neural network classifier head. 
We use the 125M parameter model architecture of Microsoft's CodeBERT~\cite{feng2020codebert}, which itself is based on BERT~\cite{devlin2018bert} and RoBERTa~\cite{liu2019roberta}. 
We elide details of the Transformer architecture and the associated training methodology, which can be found in the relevant works~\cite{devlin2018bert, feng2020codebert, liu2019roberta}. 
The Code Classifier additionally has a sequence classification/regression head that uses the same architecture and training methodology as the RoBERTa model for sequence classification~\cite{liu2019roberta}.

The Code Classifier is trained using 90\% of the dataset gathered in Section~\ref{hsr1} and is evaluated on accuracy and f1-score using the randomly-selected, held-out 10\% of the dataset. 
We pretrained on the Masked Language Modeling (MLM) task to improve syntactic understanding of the domain. 
This is the same method used to train CodeBERT on the MLM task~\cite{feng2020codebert}. 
This improves the feature vector intermediate representation generated by the base Transformer. 
This vector is used by the classification head to represent the syntax of the input sequence in vector form, which can be classified using more classic deep learning techniques~\cite{farahani2021brief}. 
We perform this step to improve the model's understanding of C/C++ and of smaller code Snippets within functions. 
This is necessary because CodeBERT is trained on only entire functions and their associated comments in a variety of languages not including C/C++. 
While CodeBERT generalizes to out-of-scope languages~\cite{feng2020codebert}, the model could be misled by certain language idiosyncrasies like whitespace in Python or naming conventions in Java. 
This step results in a 24\% increase in accuracy and f1-score.

Next, we train the Code Classifier model on our classified dataset, using code (with comments removed) to determine the class. 
The Code Classifier only classifies code into one of the first four types, because these are the only types that assist with comprehension as found by HSR1 and as described in Section~\ref{prompting}. 
The Code Classifier does not include many of our comment types because its purpose is to recommend a comment type to describe a given code Snippet, and we do not want to recommend adding, for example, a TODO comment, as such a comment would not aid comprehension.

We calculated accuracy and f1-scores resulting from this training by using 10\% of the dataset held-out as testing data and taking the median of 5 trials, as with the evaluation of RoBERTa for the classification task~\cite{liu2019roberta}.
We find that the Code Classifier model achieves an accuracy and f1-score 0.96 for this task. This shows that our models can classify comments effectively in spite of our observation that human-written comments often exhibit the qualities of multiple categories simultaneously. 
This enables the Code Classifier to effectively determine which comment template is to be used to describe a given code Snippet within the \textsc{ComCat} pipeline, enabling dynamic code Snippet selection for more fine-grained comment generation and improved readability.

\subsection{Template Catalog}
The Template Catalog in this pipeline is a fixed mapping of comment types to comment templates. 
These templates detail the structure of their associated comment type for ChatGPT using the Template prompt pattern~\cite{white2023prompt} and they are designed to answer questions programmers ask during software evaluation tasks~\cite{sillito_2006, hu2022practitioners} (Section~\ref{prompting}). 
The Template Catalog consists of four comment templates corresponding to each of the four classes that could be output by the Code Classifier. 
The purpose of this component is to maintain a set of templates that can be incorporated into prompts to outline the structure of the comment that we want to generate for a specific Snippet of code. 
This component is queried by the prompter using the class of the current code Snippet as the key and giving the associated template as the associated value.

\subsection{Template-Based Prompter}
The Template-Based Prompter constructs a list of prompts that each requests comments of a particular format to describe the code. 
It takes a list of code Snippets and corresponding comment type classifications as input and outputs a list of prompts. The prompter takes each classified code Snippet, queries the Template Catalog to retrieve the comment template associated with code Snippet's class, and crafts a prompt consisting of ChatGPT's instructions, the template, and the code Snippet. These instructions detail ChatGPT's task to generate a comment, specify output that describes the code's functionality and purpose, and explain how to use the template. 
We further limit inline comment length in these prompts to 30 words following previous work suggesting that comments over 30 words should be reserved for descriptions of entire functions or files~\cite{steidl_2013}. The prompter is important because careful prompt engineering is critical to contextualizing requests and ensuring specific qualities of LLM outputs~\cite{white2023prompt}.

\subsection{ChatGPT} 
To produce comments describing code snippets from an input file, we query ChatGPT with prompts output by the prompter. 
ChatGPT is a state-of-the-art LLM capable of many tasks, including code summarization~\cite{chatgpt}. 
It enables us to specify a complex output while providing context for the input code to produce an accurate and comprehensible comment.

We interact with ChatGPT for a single conversation for the process of commenting a file. 
We first use a system prompt to guide the model's behavior throughout the conversation. 
Our system prompt specifies ChatGPT's role as a software comment generator and provides the entire file to be commented. 
We give ChatGPT the entire file prior to the individual code Snippets so that it has context for those Snippets within the file that it will comment, as context has been shown to improve code comment generation~\cite{medhat2014sentiment,phan2021cotext,terry2020thinking}.

Following the system prompt, we use user prompts to generate comments for individual code Snippets. 
User prompts are intended for standard conversational interactions with ChatGPT, and we use them to make requests with prompts generated by the Prompter. 
These prompts are fed to ChatGPT in sequence and a comment is generated for each prompt. 
This generates every comment to be included in the newly-commented file. 
Comments are then placed on the line above the associated code Snippet in the original file. 
Finally, the complete file with embedded comments is output to the user.

\section{Evaluation and Results}
\label{eval}
In this section, we evaluate \textsc{ComCat}'s effectiveness at producing accurate and readable comments. 
We conducted two additional HSR studies to evaluate our pipeline according to (1) developer performance and (2) developer preference. We evaluate several research questions based on our findings:

\squishlist
    \item[\bfseries RQ1] Does \textsc{ComCat} lead to improved developer performance on code comprehension tasks?
    \item[\bfseries RQ2] Which code comprehension tasks are most aided by \textsc{ComCat}?
    \item[\bfseries RQ3] Do developers perceive \textsc{ComCat} comments as higher quality than human-written and standard ChatGPT-generated comments?
\squishend

We present the results for each of these research questions and discuss their interpretations.
All participants in these studies were recruited from student and professional populations of developers with at least 2 years of C/C++ experience from a variety of universities and companies through email solicitations and online crowdsourcing. 
All participants must pass a C/C++ code comprehension prescreening survey.
This survey is administered online via a URL given to participants.

\subsection{RQ1: Developer Performance}
\label{rq1}
For this research question, we seek to understand if \textsc{ComCat} leads to objectively improved developer performance on code comprehension tasks. 
We evaluate developer performance on such tasks given code that has been commented by humans and \textsc{ComCat}. 
To that end, we conduct a study of 30 developers (15 students and 15 professionals) to complete a set of indicative code comprehension and debugging tasks.
This is a 16 question survey; each question consists of a syntax-highlighted code window and a question drawn from three types: short answer, code writing, or debugging. 
We selected code for each question from our dataset (Section~\ref{hsr1}) that were between 15 and 50 lines long and contained at least one of the four comment types from our schema.
We further validate that each code snippet does not require external information (e.g., a reference to an external file) about its project to answer the questions we ask. 
Questions cover a variety of typical software engineering topics, including string manipulation, data structures, memory management, and concurrency. 
For each question, each participant is randomly assigned either the \textsc{ComCat}-commented or human-commented variant of the code. 
The code is identical between cohorts; only the comment changes.
The human-written comments are those written in the original GitHub repositories from which we drew the code in our dataset --- we consider these to be ground truth, high quality human-written comments. 
With this design, we seek to evaluate whether the comments produced by our approach lead developers to be more productive or better comprehend source code than with baseline human-written comments of the same code.
Below, we discuss each type of task given to participants in this study. 

\textbf{Short Answer} questions are open-ended questions intended to assess a participant's ability to comprehend the code. 
Each participant is given 10 code files, one at a time, and is asked to answer three short answer questions per file. These questions are taken from those found to be relevant to developer comprehension by Sillito et al.~\cite{sillito_2006} and are adapted to the specific file shown to the participant. 
This follows methodology from previous work~\cite{stapleton2020human}.
Specifically, we adapted the following questions from Sillito et al.:

\squishlist
    \item Q1. Under what circumstances is this function called?
    \item Q2. How can we tell that this function has executed correctly?
    \item Q3. What will be the direct impact of this change?
\squishend

These questions were chosen because they relate to the comprehension of single functions; we did not consider questions that relate to entire classes or software ecosystems.

\textbf{Code Writing} questions involve writing a simple function in the same namespace as the given code that accomplishes some given task related to that code. 
The participant gives their answer in a syntax highlighted editor.
In each of the three questions of this type, the task required either a simple application of the given code that could be accomplished in 5 LOC or less, or it required reusing logic from the given code. 
Here, we test a participant's practical understanding of the code resulting from the comments, independent of their ability to implement complex logic.

\textbf{Debugging} questions require participants to debug a given code sample. 
They are given a textual description of the bug that explains the problem and how it has affected the output of the function. 
The code is then presented in an syntax-highlighted editor for the participant to modify.
The bugs were injected into this code by making a single line edit, ensuring a simple correct answer. 
Here we measure comprehension of the code as a result of the comments, independent of a participant's ability to implement complex logic.

Since participants could provide open-ended answers, we developed a rubric to grade each snippet-question pair to provide a quantitative assessment of participant data.
For each question, we mark an answer as correct or incorrect based on the rubric. 
This rubric gives latitude for minor mistakes unrelated to comprehension of the snippet in the case of code writing and debugging questions.
Three raters graded all responses in this study and we achieved high inter-rater agreement~\cite{fleiss1971measuring} for our binary correctness ratings ($\kappa$=0.81, ``high agreement''). 
Disagreements in correctness were resolved through discussions and majority results.

We divide our data into two groups: responses given \textsc{ComCat}-generated comments, and responses given human-written comments. 
We consider rater-assessed correctness for each response and compare the two populations of responses. 
We note a significant difference in correctness between these two populations (p$<$0.001), finding that correctness improves by 12\% given \textsc{ComCat}-generated comments over the human-written alternative. We further find that 83.33\% of participants perform better given \textsc{ComCat}-generated comments and an additional 6.33\% perform equally well between the two versions.
Participants given human-written
comments scored an average of 71.90\% correct, while those given
\textsc{ComCat}-generated comments scored an average of 83.88\% correct. 

\noindent\fbox{
    \parbox{0.97\linewidth}{
        Therefore, we find that both professional and student developers find code easier to comprehend when \textsc{ComCat} documents the code compared to developer-written documentation. Thus, our expertise-guided context can improve automated comment generation to aid developers in comprehension and maintenance.
    }
}

\begin{figure}
    \centering
        \includegraphics[width=\linewidth]{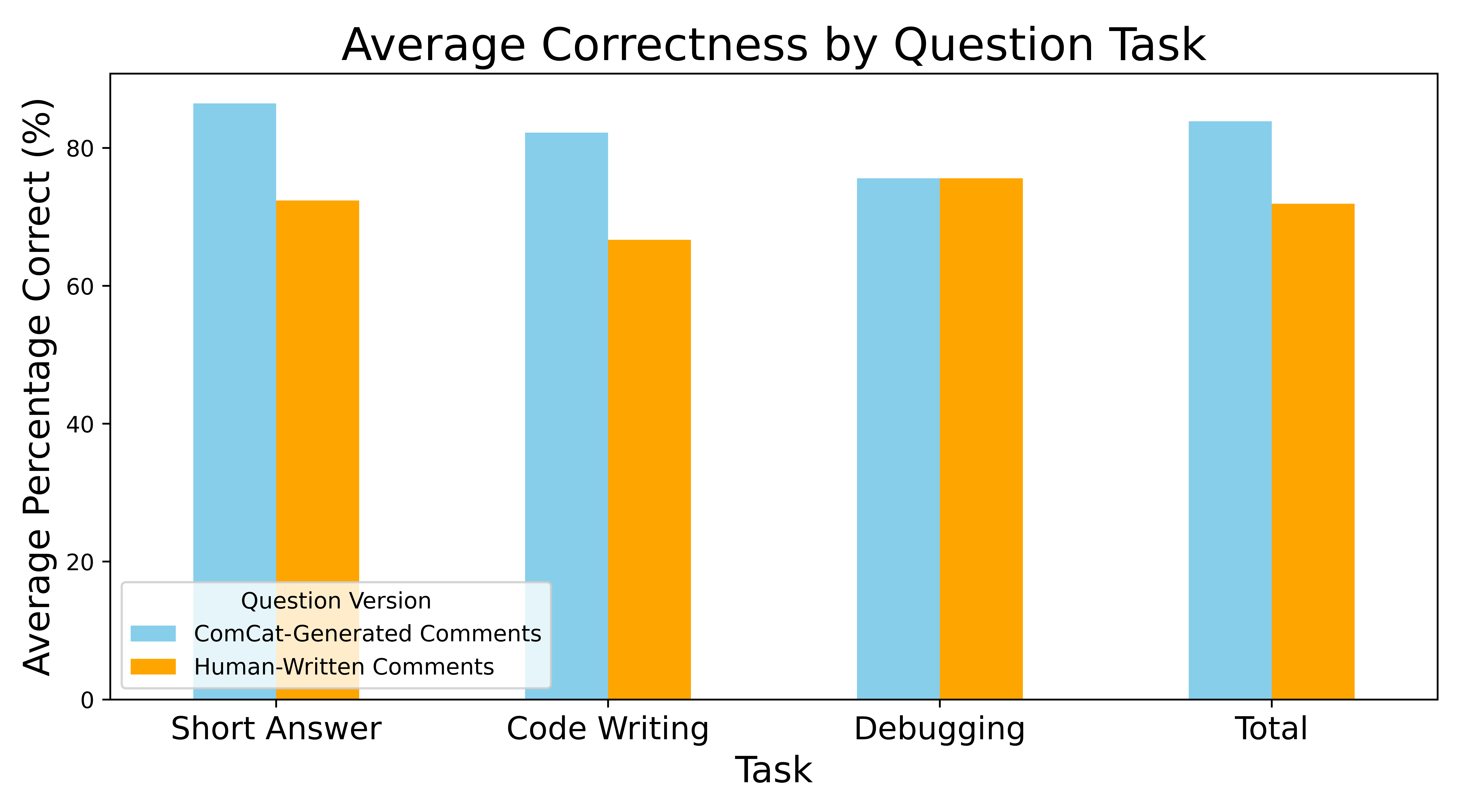} 
    \vspace{-4ex}
    \caption{Results from HSR2. This shows the difference in correctness between \textsc{ComCat}-generated and human-written comments for each question type.}
    \label{fig:correctness}
    \vspace{-3ex}
\end{figure}

\subsection{RQ2: \textsc{ComCat} Performance on Varied Tasks}
\label{rq2}
Following our evaluation of \textsc{ComCat}'s effect on developer performance on code comprehension tasks, we further investigate performance on individual comprehension tasks. Here we evaluate participants in our human performance study (cf. Section~\ref{rq1}) on 
our three question types
(i.e. short answer, code writing, and debugging). We again measure the change in correctness between responses resulting from \textsc{ComCat}-generated versus human-written comments, but we isolate each task to determine how performance on that task is affected by our intervention. The results from this investigation, as well as the overall change, are summarized in Figure~\ref{fig:correctness}.

\subsubsection{Short Answer Task}
This task involved answering open-ended questions found by Sillito et al.~\cite{sillito_2006} to be important questions for comprehension that are commonly asked of developers by project leads. 
We find that participants answer these questions correctly significantly more often given \textsc{ComCat}-generated comments (p$<$0.001). 
More precisely, the correctness is improved by 14.08\% from 72.38\% correct (for human-written) to 86.46\% correct (for \textsc{ComCat}). 
Furthermore, 90\% of participants performed better on this tasks using \textsc{ComCat} as well. 
This indicates that \textsc{ComCat} outperforms human-written comments in helping developers answer project-relevant questions. We attribute this to inline comments assisting developers in answering more focused comprehension questions.

\subsubsection{Code Writing Task}
This task involved solving a short coding problem by using given functions based on documentation for those functions. We find that participants answer questions correctly more often given \textsc{ComCat}-generated comments; however, the difference is not significant (p=0.097). 
Here, correctness is improved by 15.55\% from 66.67\% correct (for human-written) to 82.22\% correct (for \textsc{ComCat}). 
While these results do not reflect significant improvements in code writing, these results do at least suggest that \textsc{ComCat} generates comments that are no worse than ground-truth comments written by developers.
We attribute this to our function-type comments aiding developers in function interface use by describing parameters and return values, and our inline comment types further assist developers in reusing the logical intuition from Snippets within the code. 

\subsubsection{Debugging Task}
This task requires participants to debug a code snippet given a natural language description of the buggy behavior. This requires participants to identify and fix a single buggy line of code. 
We find that, on average, participants performed exactly the same given both versions of the commented code, so we cannot make any conclusions regarding the superiority of either version in aiding participants for this task. Specifically, participants averaged 75.6\% correctness when given both the \textsc{ComCat}-generated comments and the human-written comments. 
Thus, we can only surmise that \textsc{ComCat}-generated comments are likely no worse than human-written comments. We attribute this to \textsc{ComCat}-generated comments describing what the code is actually doing, not what it is intended to do. 
This could explain why we do not see improved correctness for this task.

Overall, our investigation into \textsc{ComCat}'s performance across varied comprehension tasks reveals nuanced insights into its effectiveness. 
The short answer task showed significant results in \textsc{ComCat}'s performance versus that of human-written comments, indicating improved comprehension.
Meanwhile ComCat provided equally helpful documentation for code writing and debugging tasks.

\noindent\fbox{
    \parbox{0.97\linewidth}{
        \textsc{ComCat} can improve developer correctness on project-relevant questions without detrimentally impacting debugging or code reuse. 
        This is an improvement over previous function-level summarization, as such comments are less effective than human-written comments~\cite{stapleton2020human}.
    }
}

\subsection{RQ3: Developer Preference}
\label{rq3}
For this research question, we evaluate developer perceptions of \textsc{ComCat}'s comments. 
We assess developer preference for \textsc{ComCat} versus human-written comments and standard ChatGPT through two main evaluation tasks: describing individual code Snippets and automatically commenting files. 
We designed another human study of 32 developers focusing on their comment preferences. 
We do not compare \textsc{ComCat}'s performance with models specifically trained for code summarization in C/C++ because such models are only evaluated on summaries of entire functions and are incapable of parsing files for code Snippets to summarize (e.g., specific declarations, statements, or loop constructs). 
Moreover, these models are trained to replicate ground truth human-generated comments, which we include in this study anyway.
We choose ChatGPT for our LLM comparison rather than other LLMs, like Google's Bard, because its architecture is the state-of-the-art in code summarization~\cite{khan2022automatic} and is the most widely used for AI programming assistance through GitHub Copilot (based on OpenAI's Codex)~\cite{copilot, khan2022automatic}.
Therefore, we compare to human-generated comments and ChatGPT, which we observe might produce overly verbose documentation that might not aid in comprehension, as we show with this study.

\begin{figure}
    \centering
        \includegraphics[width=\linewidth]{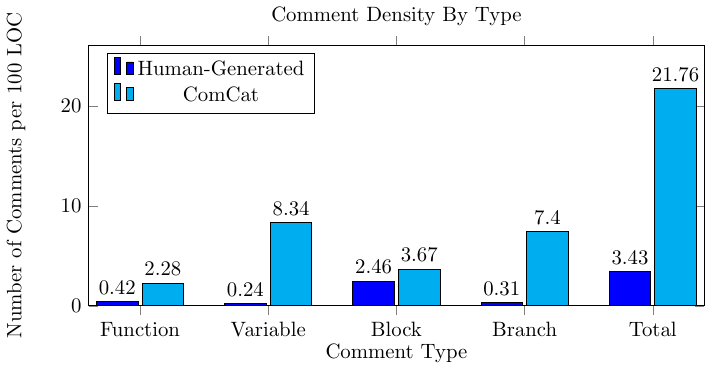} 
    \vspace{-5ex}
    \caption{Density of comment types in our dataset. Here, density is the number of comments per 100 LOC. \textsc{ComCat} improves comment density over humans, addressing the problem of sparsely commented, minimally maintainable code without cluttering code with excessive documentation.}
    \label{fig:bar}
    \vspace{-5ex}
\end{figure}

Commonly, software that is documented by humans is sparsely-commented and existing research lists comment completeness and missing or insufficient comments as major issues with existing documentation~\cite{aghajani2020software, plosch2014value, steidl_2013, xia2017measuring}.
The goal of this pipeline is to automatically improve code readability and maintainability by generating comments of various types.
This requires a balance in comment density to ensure the documentation is complete, but is not so verbose as to clutter the code.
We illustrate this problem in Figure~\ref{fig:bar}, showing that \textsc{ComCat} generates a greater density of comments for more complete documentation than what is produced by human developers\footnote{This data was gathered from our dataset, described in Section~\ref{classification}. Every file contained at least one human-written comment.}.
Figure~\ref{fig:bar} further shows that we generate a range of comment types more completely than humans, and humans tend to produce very few comments that explain variables or branches.
\textsc{ComCat} further documents every instance of a function or method with a function comment, where humans in this data documented only 18.42\% of functions.
We similarly calculated these metrics for comments produced by standard ChatGPT. 
We found that while it documents nearly every function with a function comment, it rarely produces variable and branch comments, potentially neglecting important semantics for comprehension.
We show in our experiments that \textsc{ComCat}'s comments are subjectively individually as good as those generated by humans and do not clutter code with excessive documentation, as evidenced by developer preference for comments, striking the balance between completeness and succinctness that we desire.
The following subsections describe this evaluation, indicating that \textsc{ComCat} can improve documentation through increased completeness without sacrificing individual comment quality.

\subsubsection{Experimental Design}
\label{exp_design}
For the first preference evaluation task in this study, we assess \textsc{ComCat}'s ability to generate comments that accurately and comprehensively describe relevant information about individual code Snippets within source code according to developer preferences. 
This shows that \textsc{ComCat}'s comments are individually at least as preferred as those generated by humans and standard ChatGPT. 
We randomly select code Snippets and their associated human-generated comments from the held-out test data (Section~\ref{model}) in our dataset (Section~\ref{hsr1}). 
However, we ensure that the human-written comments from the dataset are one of the four types of comments we established in our schema.
Because the code Snippets are already chosen, this task does not evaluate the Code Parser; instead, single code Snippets are passed on to the next stage of the pipeline. 
Testing against human-written comments allows us to explore the pipeline's usefulness in supplementing or supplanting manual commenting. 
Here we demonstrate \textsc{ComCat}'s ability to save developer effort by automatically commenting code Snippets in a way that is at least as accurate and readable as human-generated comments. 
In separate trials for the same task, using the same code Snippets, we compare to comments produced by standard ChatGPT to evaluate our pipeline against a state-of-the-art LLM, demonstrating the effectiveness of our model augmentation and capturing improvements to ChatGPT’s comment generation. For this comparison, we prompt ChatGPT as similarly as possible to \textsc{ComCat} without the assistance of our pipeline to mitigate prompt selection biases. 

For our second preference evaluation task, we assess our pipeline's ability to comment entire C/C++ files and to see whether developers prefer these comments over other automated comment generation techniques and human-written comments. 
We use a diverse set of source files, covering various software domains and complexity levels, drawn randomly from our dataset. 
The C/C++ files are fed into our pipeline, and the output is compared to comments written by the code author and generated by standard ChatGPT. 
We select code Snippets and human-generated comments in the same way as in the evaluation task for individual comments and we again prompt standard ChatGPT as similarly as possible to \textsc{ComCat}. 

HSR3 is a survey of 32 developers, who provide subjective feedback on the quality and relevance of the generated comments compared to human-generated and Standard ChatGPT-generated comments.
Every participant in this study answered every question for the trial they were given.
For each question in this study, participants are presented two windows containing C/C++ code and associated comment(s), where the code is identical in both windows, but the comment(s) are generated differently. 
Participants are not informed of the method used to generate each comment.
We ask participants to indicate a preference (or no preference) between the two alternatives for which example they find to be more ``readable, accurate, and useful for understanding, editing, debugging, or reusing the associated code.''
At the end of each trial, participants are asked to explain their preferences and describe qualities they tended to like or dislike in each option.
Participants choose between \textsc{ComCat}'s comment and either one produced by Standard ChatGPT or a human-written comment from our dataset. 
This design allows us to garner specific, comparative insights into the utility and quality of the ComCat-generated comments according to developer preference, aiming to ascertain developers' subjective preferences about our pipeline in enhancing code comprehensibility.

\begin{table}
    \caption{Results from our developer preference survey. This shows the percent of questions for which a majority of participants indicated preference for \textsc{ComCat} over comments generated by humans and standard ChatGPT. \vspace{-5pt}}
    \label{table:preference}
    \begin{tabular}{ |p{3cm}|p{2cm}|p{2cm}| }
     \hline
     \multicolumn{3}{|c|}{Developer Preference for \textsc{ComCat}} \\
     \hline
     \rule{0pt}{3ex} 
     Task & vs. Humans & vs. ChatGPT
     \rule[-1.5ex]{0pt}{0pt}
     \\
     \hline
     \rule{0pt}{3ex} 
     Individual Comments & 
     \hspace{20pt}\textbf{80\%} &
     \hspace{20pt}\textbf{92\%} 
     \rule[-1.5ex]{0pt}{0pt}
     \\
     \hline
     \rule{0pt}{3ex} 
     Entire Files &
     \hspace{18pt}\textbf{100\%} &
     \hspace{20pt}\textbf{60\%} 
     \rule[-1.5ex]{0pt}{0pt}
     \\
     \hline
    \end{tabular}
    \vspace{-10pt}
\end{table}

\begin{table}
    \caption{Further results from our survey. This shows the average percent of times each participant preferred \textsc{ComCat} over comments generated by humans and standard ChatGPT. \vspace{-5pt}}
    \label{table:preference-part}
    \begin{tabular}{ |p{3cm}|p{2cm}|p{2cm}| }
     \hline
     \multicolumn{3}{|c|}{Developer Preference for \textsc{ComCat}} \\
     \hline
     \rule{0pt}{3ex} 
     Task & vs. Humans & vs. ChatGPT
     \rule[-1.5ex]{0pt}{0pt}
     \\
     \hline
     \rule{0pt}{3ex} 
     Individual Comments & 
     \textbf{66\%} \hspace{2pt} (p$<$0.001)&
     \textbf{78\%} \hspace{2pt} (p$<$0.001)
     \rule[-1.5ex]{0pt}{0pt}
     \vspace{-2ex}
     \\
     \hline
     \rule{0pt}{3ex} 
     Entire Files &
     \textbf{82\%} \hspace{2pt} (p$<$0.001)&
     \textbf{56\%} \hspace{2pt} (p=0.082)
     \rule[-1.5ex]{0pt}{0pt}
     \\
     \hline
    \end{tabular}
    \vspace{-15pt}
\end{table}

\subsubsection{Preference Study Results}

Here, we discuss the results of HSR3, which shows how \textsc{ComCat} aligns with developers' expectations. 
A summary of our findings can be found in Tables~\ref{table:preference}~and~\ref{table:preference-part}. 
We measure the proportion of questions for which a majority of participants indicated preference for \textsc{ComCat} in each of our trials (c.f. Table~\ref{table:preference}), along with the mean percent of questions that each participant answered with preference for \textsc{ComCat} (c.f. Table~\ref{table:preference-part}). 
Higher or equal preference for comments generated by our pipeline over human-generated comments may suggest \textsc{ComCat}'s comments could replace or augment human commenters in such situations. 
Higher preference for \textsc{ComCat} over standard ChatGPT indicates improvements over the state-of-the-art and showcases the subjective effectiveness of our pipeline's context. 
Table~\ref{table:preference-part} shows results calculated by participant. 
In this Table, we calculate p-values for each trial and find all trials to be statistically significant, except for the instance where we comment entire files and compare to standard ChatGPT, which remains inconclusive. 
However, we can conclude that \textsc{ComCat} is no worse than standard ChatGPT in this case.

Our results indicate that the \textsc{ComCat} pipeline is effective at generating informative and relevant comments according to the standards of experienced developers.
We find that for non-trivial input code Snippets, \textsc{ComCat}'s comments are at least as preferred as human-generated comments in 80\% of cases individually, and in 100\% of cases for entire files where humans tend to under-comment. 
\textsc{ComCat} further produces comments that are preferred to those produced by standard ChatGPT in 92\% of cases for individual comments and 60\% of cases for entire files according to our results. 

\noindent\fbox{
    \parbox{0.97\linewidth}{
        Developers find \textsc{ComCat}'s comments to be more useful for code reading compared to human-written and standard ChatGPT-generated comments, showcasing the pipeline's practical benefits in assisting with code comprehension and documentation tasks.
    }
}

\section{Discussion}

Our results demonstrate \textsc{ComCat}'s effectiveness in automating comment generation, showing that our pipeline can improve developer comprehension of source code. 
In this study, we found that professional and student developers find code easier to comprehend when \textsc{ComCat} documents the code compared to when developers write documentation. 
Participants also prefer \textsc{ComCat} to human-written documentation in subjective comparisons. 
We intuit that this is due to appropriate comment structure and density, where human-written documentation can be sparse and is often ``written only for the author's understanding,'' as participants note in HSR3. 

In addition to the analyses described in Section~\ref{eval}, we considered several other potential hypotheses that did not yield significant results. 
We examined expertise (measured by years of programming experience) within our participant cohorts in HSR2, and found no significant difference in the change in correctness when given \textsc{ComCat}-generated versus human-written comments (p=0.21). 
Our current results suggest that \textsc{ComCat} is similarly helpful for both students and professionals (i.e., novices and experts). 
Next, we also recorded and analyzed the time taken to complete each question for each participant in HSR2, but found no significant change in time taken (p=0.17). 
Thus, we conclude that \textsc{ComCat} has no significant effect on the time taken to comprehend code when compared to code that is commented by humans.

\subsection{Threats to Validity}
While our evaluation provides key insights, there are potential threats to the validity of our approach. First, our dataset of C/C++ code and comment types used for training and evaluation might have biases inherent in the code and comments scraped from GitHub projects. 
We further use subjective human raters for this data to determine comment types, developer preferences, and participant correctness. 
We mitigate these threats by randomly choosing code Snippets from well-maintained repositories and we prescreen participants for C/C++ aptitude and experience. 
We further calculate inter-rater reliability metrics to ensure sound agreement.

Furthermore, while we focus on comment generation in C/C++, we expect that the insights derived from this research are applicable to other programming languages due to previous code summarization work that is applicable across languages and the generality of ChatGPT~\cite{ahmad2021unified, bang2023multitask, feng2020codebert, husain2019codesearchnet, khan2022automatic, phan2021cotext, wang2021codet5}. Our evaluation of ChatGPT is also reliant on prompt quality. To ensure a fair comparison, we use best practices for prompt engineering~\cite{white2023prompt} and we prompt standard ChatGPT as similarly as possible to \textsc{ComCat}'s variant.

\section{Conclusion}
In this paper, we present the \textsc{ComCat} pipeline, which uses expertise-guided context generation from human subjects research to automate the generation of inline and function-level comments. 
Our evaluation demonstrates the effectiveness of this context generation in enhancing code readability and comprehension, surpassing human-generated comments' ability to assist developers in answering project-relevant questions. 
By following developer preferences for comment types and structures, \textsc{ComCat} generates informative and relevant documentation that aids developers in understanding software. 
Ultimately, \textsc{ComCat} represents a step towards more effective software maintenance and documentation, offering tangible benefits on the lifetime costs of software, and serving as a compelling solution for modern software engineering problems.

\bibliographystyle{IEEEtran}
\bibliography{IEEErefs}

\end{document}